# FAST: A **F**uture **A**ircraft **S**izing **T**ool for Advanced Aircraft and Propulsion System Design

Paul R. Mokotoff ⬥ [1], Maxfield Arnson ⬥ [1], and Gokcin Cinar ⬥ [1]¶

**1** Department of Aerospace Engineering, University of Michigan ¶ Corresponding author

## Summary

ICAO predicts that, without radical technological advancements, the global aviation industry will emit up to 28 gigatons of $CO_2$ between 2020 and 2050 (International Civil Aviation Organization, 2022). To reduce aviation-related emissions, innovative aircraft technology, including electrified aircraft propulsion, is under development. For instance, NASA's Electrified Powertrain Flight Demonstration (EPFD) Project is advancing these technologies with U.S. industry partners (National Aeronautics and Space Administration, 2022). However, current aircraft sizing tools require detailed design information that may not be available early in the development process, particularly for novel technologies. This can yield sub-optimal designs and inhibits innovation. Thus, a computational tool is needed to easily and rapidly size an aircraft configuration while allowing the designer to explore the design space, examine tradeoffs, and evaluate alternative designs.

The **F**uture **A**ircraft **S**izing **T**ool (**FAST**) is a Matlab-based tool that addresses this challenge by rapidly sizing aircraft with *any* propulsion architecture, including conventional, electric, and hybrid-electric systems, even with limited initial data. FAST enables engineers to explore various aircraft configurations, evaluate design alternatives, assess performance across a flight envelope, and visualize concepts during the sizing process. By supporting early-stage design, FAST addresses a gap in currently available computational tools for developing sustainable aviation technologies to help reduce the industry's carbon footprint.

## Statement of Need

During early-phase conceptual aircraft design, engineers know little information about their design. This is particularly challenging for novel aircraft designs, as key parameters may be unknown at the outset. Existing aircraft design tools are intended for late conceptual design, requiring detailed design information and a geometric configuration upfront (Cinar, 2018; David et al., 2021; Gratz et al., 2024), or are coupled with design optimization environments (David et al., 2021; Gratz et al., 2024; Lukaczyk et al., 2015) such as OpenMDAO (Gray et al., 2019). These tools can accommodate more detailed analyses, such as component weight estimations (Cinar, 2018; Gratz et al., 2024) or low-fidelity aerodynamic analyses (David et al., 2021; Lukaczyk et al., 2015). Such analyses require the user to select an aircraft configuration a priori, which prohibits rapid design space exploration.

FAST addresses these challenges by leveraging historical data from over 450 aircraft and 200 engines to create predictive regressions (M. Arnson et al., 2025). These regressions are employed to estimate any design parameters that the user may not know, such as engine weight based on required thrust or power. Combined with physics-based models, FAST rapidly analyzes an aircraft configuration, converging on a design faster than existing detailed design tools. This rapid sizing is achieved through an energy-based mission analysis, which approximates the aircraft as a point mass to evaluate the aircraft dynamics, forces, and energy required for a given mission (Anderson, 1999; Cinar, 2018). Additionally, FAST can analyze any propulsion architecture, including conventional, electric, and hybrid systems, using a graph theory-inspired approach that represents propulsion system connections and in-flight operations as matrices (Cinar et al., 2020).

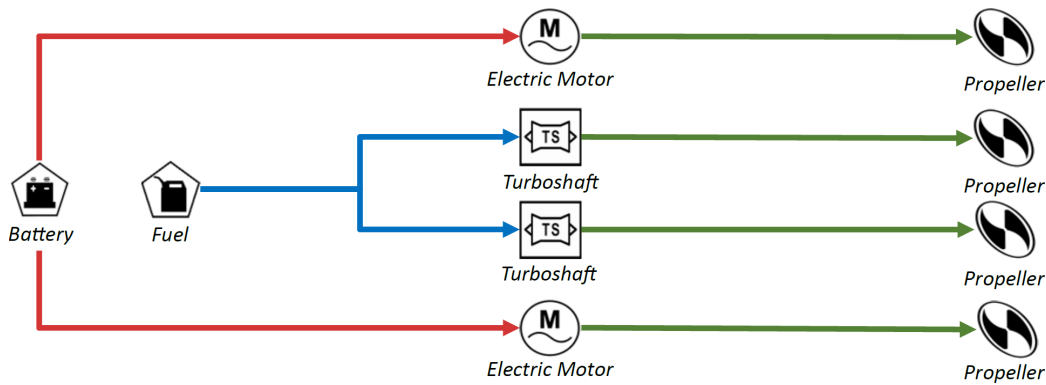

**Figure 1:** Electrified freighter propulsion architecture.

FAST has been utilized in multiple research projects within NASA's EPFD project to assess the performance of conventional and electrified aircraft. First, a commercial freighter was electrified by replacing the outboard gas-turbine engines with electric motors (Mokotoff et al., 2025), shown in Figure 1. The trade studies explored the fuel burn savings achieved by removing a fraction of the freighter's payload to accommodate the electrified propulsion system components. Also, NASA's **SU**bsonic **S**ingle **A**ft E**n**gine (**SUSAN**) concept was modeled in FAST and explored how advanced technologies impacted the aircraft's performance (Wang et al., 2025). More granular aerodynamic and propulsive models were incorporated into FAST to better assess the benefits of boundary layer ingestion and distributed electric propulsion technologies.

FAST was also used to explore advanced propulsion system designs and their impacts on the final aircraft design (M. G. Arnson et al., 2024). Advanced ATR 42-600 configurations were designed with battery electric and hydrogen fuel cell electric propulsion systems. The aircraft's performance was assessed while varying key performance parameters such as the battery gravimetric specific energy, powertrain efficiency, and fuel cell power-to-weight ratio.

ARPA-E used FAST to estimate the fuel/operating costs for electrified aircraft on domestic US flights (de Bock, 2024). Also, a fleet of hybrid electric aircraft was sized and operated on routes currently flown by regional jets using FAST (Deng et al., 2023). This work demonstrated how hybrid electric aircraft can be integrated into a regional airline's fleet while maintaining the same operational capabilities as existing aircraft, assuming that the necessary battery-charging infrastructure and power supplies are available.

# FAST Workflow Overview

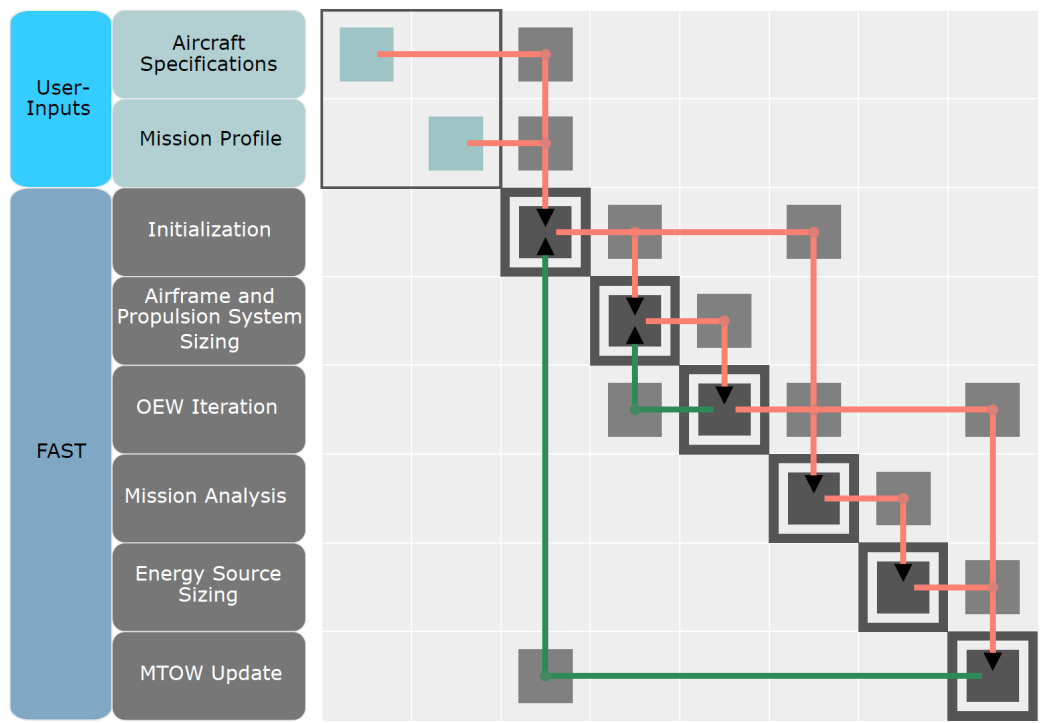

**Figure 2:** High-level overview of FAST's main functionality, produced using (Gray et al., 2019).

Figure 2 provides an overview of FAST's functionality, detailing all user inputs and modules. Data is passed forwards (colored red) and provided as feedback (colored green) in an iterative process.

First, the user provides the Aircraft Specifications and a Mission Profile, which informs FAST how to fly the aircraft while sizing it. Then, FAST's Initialization module assembles the user-provided information into an aircraft model and utilizes its historical database (M. Arnson et al., 2025) to generate regressions and predict any unknown parameters. Once the initial model is complete, FAST generates mathematical representations of the propulsion architecture and its operation during flight (Cinar et al., 2020).

The aircraft is sized using a fixed-point iteration (Ascher & Greif, 2011). First, an inner iteration sizes the airframe and propulsion system between the "Airframe and Propulsion System Sizing" and "Weight Build-Up" modules. FAST does not perform a constraint analysis and assumes that the thrust- or power-weight ratio and wing loading provided remain fixed and are feasible. Then, an energy-based mission analysis (Anderson, 1999; Cinar, 2018) calculates the energy required for the mission and allocates it amongst the available energy sources (jet fuel, hydrogen, battery). Lastly, the energy required from the mission analysis informs the "Energy Source Sizing", which updates the aircraft's weight. The iteration continues until converging on a sized aircraft.

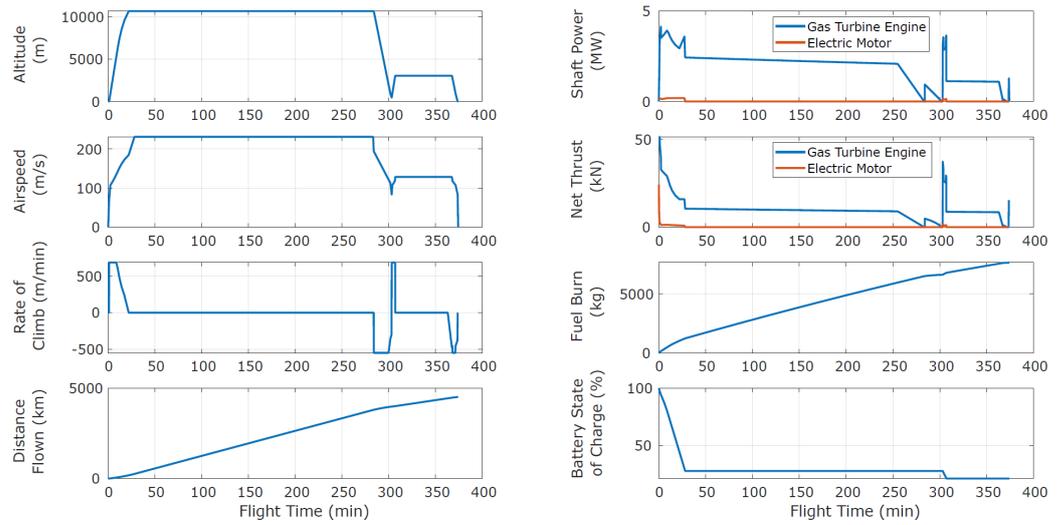

Figure 3: Example mission history plots.

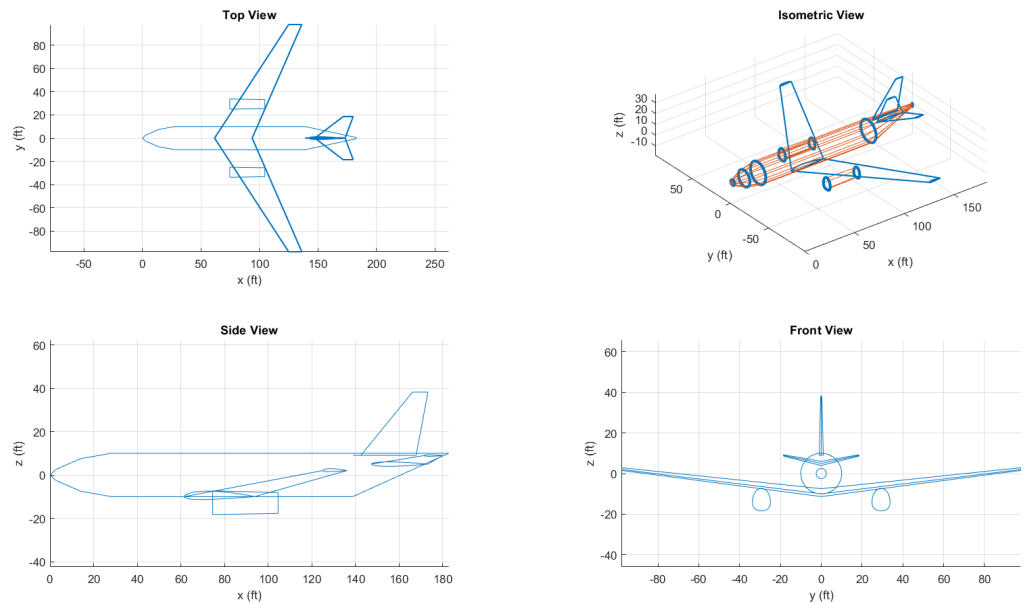

Figure 4: Transport aircraft geometry shipped with FAST.

After sizing, the aircraft model is returned as a data structure, allowing for further analysis or integration into other studies. FAST also offers post-processing options, including mission history visualization (see Figure 3) and geometric visualization of the sized aircraft (see Figure 4). To visualize an aircraft concept, users either prescribe their own aircraft geometry or use one that is pre-defined within FAST (Khailany et al., 2025).

# Acknowledgements

This work is sponsored by the NASA Aeronautics Research Mission Directorate and the Electrified Powertrain Flight Demonstration (EPFD) project, "Development of a Parametrically Driven Electrified Aircraft Design and Optimization Tool". The IDEAS Lab would like to thank Ralph Jansen, Andrew Meade, Karin Bozak, Amy Chicatelli, Noah Listgarten, Dennis

Rohn, and Gaudy Bezos-O'Connor from the NASA EPFD project for supporting this work and providing valuable technical input and feedback throughout the duration of the project. The work was performed under Glenn Engineering and Research Support Contract (GEARS) Contract No. 80GRC020D0003.

The authors would also like to thank Huseyin Acar, Nawa Khailany, Janki Patel, and Michael Tsai for their contributions to developing FAST.